\documentclass[12pt,titlepage]{article}
\usepackage{amsmath}
\usepackage{amssymb}
\usepackage{graphicx}
\usepackage{caption2}
\usepackage{amsfonts}
\usepackage{cite}
\usepackage{lineno}

\newcommand{\be}{\begin{equation}}
\newcommand{\ee}{\end{equation}}
\newcommand{\bea}{\begin{eqnarray}}
\newcommand{\eea}{\end{eqnarray}}
\newcommand{\bef}{\begin{figure}}
\newcommand{\ef}{\end{figure}}
\newcommand{\bt}{\begin{tabular}}
\newcommand{\et}{\end{tabular}}
\newcommand{\bno}{\begin{enumerate}}
\newcommand{\eno}{\end{enumerate}}
\newcommand{\nn}{\nonumber}

\def\3{\ss}

\catcode`\"=\active
\def"{\accent'177}

\begin{document}

\begin{center}
{\large\bf  A new non-perturbative approach in quantum mechanics \\
for time-independent Schr\"{o}dinger equations}

Shijun Liao $^{1,2,3}$\\

$^1$ School of Physics and Astronomy, Shanghai Jiao Tong University, China \\

$^2$ Center of Advanced Computing, School of Naval Architecture, Ocean and Civil Engineering,  Shanghai Jiao Tong University, China \\

$^3$ Ministry-of-Education Key Lab for Scientific and Engineering Computing, Shanghai 200240, China

( email address: sjliao@sjtu.edu.cn )

\end{center}

\hspace{-0.75cm}{\bf Abstract} {\em A new non-perturbative approach is proposed to solve time-independent Schr\"{o}dinger equations in  quantum  mechanics  and  chromodynamics (QCD).   It is based on the homotopy analysis method (HAM), which was developed by the author for highly nonlinear equations since 1992 and has been widely applied in many fields.   Unlike perturbative methods, this HAM-based approach has nothing to do with small/large physical parameters.  Besides,  convergent series solution can be obtained even if the disturbance is far from the known status.   A  nonlinear harmonic oscillator  is used as an example to illustrate the validity of this approach for disturbances that might be more than hundreds larger than the possible superior limit of the perturbative approach.  This HAM-based approach could provide us rigorous  theoretical  results  in  quantum  mechanics  and  chromodynamics (QCD), which can be directly compared with experimental data.   Obviously, this is of great benefit not only for improving the accuracy of experimental measurements but also for validating physical theories. }

\section{Motivation}

Perturbation methods \cite{Thouless1958, NayfehBook2000} are widely used in quantum mechanics and quantum chromodynamics (QCD)  \cite{DiracBook1958, GreinerBook2000, HamekaBook2004},  mainly because exact solutions can be gained in quite a few cases.   However, it is widely known that the perturbation methods \cite{Thouless1958, NayfehBook2000}  are valid only when a {\em small} physical parameter indeed exists, say, the disturbance (or departure) from the case with known exact solution must be tiny enough.  This limitation greatly restricts the applications of perturbation methods.  In practice, one often gives a first-order perturbation approximation and then checks whether or not it agrees with related experimental data, but {\em without} considering the convergence of the corresponding perturbative series.   Strictly speaking,  this is more or less ``phenomenological'' rather than ``ontological'', since it is not rigorous in mathematics.   If reliable convergent results in quantum mechanics and quantum chromodynamics (QCD) could be obtained,  one could directly compare them with experimental data.  Obviously, this is of great benefit not only for improving  experimental measurements but also for validating physical theories.       

To overcome the restrictions of perturbation methods, the author  \cite{liao92-phd, liao97-nlm, LiaoBook2003, LiaoBook2012, liao04-AMC, LIAO2009-CNSNS-A,  LIAO2010-CNSNS-B, LIAO2010-CNSNS-A} developed the homotopy analysis method (HAM), an analytic approximation method for highly nonlinear equations.   Unlike perturbation methods, the HAM is based on the homotopy in topology \cite{Hilton1953} and thus has nothing to do with any small physical parameters at all, and therefore  can solve nonlinear equations without small/large physical parameters.  More importantly, the HAM provides us a convenient way to guarantee the convergence of solution series by means of introducing a so-called ``convergence-control parameter'', which has no physical meanings so that we have freedom to choose a proper value for it to ensure convergence of solution series.   Besides,  the HAM provides us great freedom to choose initial guesses of unknowns, so that iteration can be used easily.  As a result, the HAM is valid for highly nonlinear equations.  
In addition, it has been proved that many traditional non-perturbative approaches, such as Adomian decomposition method (ADM) \cite{Adomian1991, AdomianBook1994},  the $\delta$-expansion method \cite{Karmishin1990, Awrejcewicz1998} and so on, are only special cases of the HAM.  Furthermore, it has been proved that even the famous Euler transform is also a special case of the HAM \cite{LIAO2010-CNSNS-B}.   In this way, nearly all restrictions of perturbation methods have been overcome by the HAM, as illustrated by its users in a wide range of fields with more than thousands related publications (for examples, please see \cite{VajraveluBook2012, Yabushita2007, Tao2007,  Liang2010,  Nassar2011, Mastroberardino2011, Sardanyes2015, VanGorder2015}).  It should be emphasized that the HAM has been successfully applied to theoretically predict the existence of the so-called steady-steady resonant gravity waves \cite{Xu2012}, which had been later experimentally confirmed \cite{Liu2015}.  The discovery of the steady-state resonant gravity waves \cite{TobischBook2015} illustrates the novelty and potential of the HAM, since  a truly new method should/must always bring us something new/different.      

Especially, the HAM works well even for problems with rather high nonlinearity.  For example, the convergent series solution of  Von K\`{a}rm\`{a}n plate under {\em arbitrary} uniform pressure (i.e., with {\em arbitrary} deformation) are obtained by means of the HAM \cite{Zhong2017SAM}, and besides it has been proved that {\em all} previous perturbative approaches for Von K\`{a}rm\`{a}n plate  are special cases of the HAM.  In addition, using the HAM,  Zhong and Liao \cite{Zhong2018JFM}  successfully gained the convergent series solution of the limiting Stokes wave of {\em extreme} height in {\em arbitrary} water depth (including the extremely shallow water),  which could not been found by perturbation methods and even by numerical techniques.  All of these illustrate that the HAM is indeed valid for highly nonlinear problems.            

Here, encouraged by all of these,  we would like to apply the HAM to quantum mechanics.   
For the sake of simplicity, let us first consider the  time-independent Schr\"{o}dinger equation
\begin{equation}
H \psi_n({\bf r})  =  E_n \psi_n({\bf r}),  \label{geq:original}
\end{equation}
where $H$ is a Hamiltonian operator, $\psi_n({\bf r})$ and $E_n$ are the unknown eigenfunction and eigenvalue of $H$, $\bf r$ denotes the spatial coordinate, respectively.  Assume that  each eigenvalue $E_n$ corresponds to an unique eigenfunction $\psi_n({\bf r})$ only.   Assume that the unknown eigenfunction $\psi_n({\bf r})$ can be expressed by a complete set of the known eigenfunctions $\psi_m^{b}({\bf r})$, $m=0,1,2,3,\cdots, N_s$, i.e. 
\begin{equation}
\psi_n({\bf r}) = \sum_{m=0}^{N_s}  a_{n,m} \; \psi_m^{b}({\bf r}) ,  
\end{equation}
 satisfying 
\begin{equation}
H_0 \psi_m^{b}({\bf r})  =  E_m^{b} \psi_m^{b}({\bf r}),  \label{geq:known}
\end{equation}
where $\psi_m^{b}({\bf r})$ and $E_m^{b}$ are the known eigenfunction and eigenvalue of  the Hamiltonian operator $H_0$, respectively, and $N_s$ should be infinite in theory but often a finite positive integer in practice.    Assume that each eigenvalue $E_m^{b}$ corresponds to an unique eigenfunction $\psi_m^{b}({\bf r})$ only, and besides  the basis $\psi_m^{b}({\bf r})$ is orthonormal, i.e. 
\begin{equation}
\int \psi_m^{b}({\bf r}) \psi_n^{b}({\bf r})^* d\Omega  =\left(\psi_m^{b}({\bf r}), \psi_n^{b}({\bf r})^*  \right) = \delta_{mn} = \left\{  
\begin{array} {l l}
1, & \mbox{when $m=n$} \\
0, & \mbox{when $m\neq n$},
\end{array}  \right.
\end{equation}
where  $ \psi_n^{b}({\bf r})^*$ is a complex conjugate of $ \psi_n^{b}({\bf r})$.  

In \S~2 we briefly describe the basic ideas of the HAM-based approach for time-independent Schr\"{o}dinger equations.  We illustrate  in \S~3 that the convergent series  can be gained by means of the HAM-based approach even in the case  far from the known situation, i.e. with quite large disturbance.   This is quite different from the perturbative approach in quantum mechanics.  For the sake of comparison,  its perturbative results are also given in \S~3.1.   The concluding remarks are given in \S~4.   For the sake of convenience, the perturbative approach is briefly described in the appendix.    

\section{The HAM-based approach}

The HAM \cite{liao92-phd, liao97-nlm, LiaoBook2003, LiaoBook2012, liao04-AMC, LIAO2009-CNSNS-A,  LIAO2010-CNSNS-B, LIAO2010-CNSNS-A} is based on the homotopy in topology \cite{Hilton1953}.   Let $q\in[0,1]$ denote the embedding parameter,  $c_0\neq 0$ be the so-called ``convergence-control parameter'', respectively.  We construct a family of equations, namely the zeroth-order deformation equation: 
\begin{eqnarray}
&& (1-q)  \left( H_0 - E_n^b\right)   \left[ \Psi_n({\bf r};q) -  \psi_n^{(0)}({\bf r})  \right]  \nn \\
&=&  c_0 \; q \; \left\{  H \Psi_n({\bf r};q)  -  {\cal E}_n(q) \Psi_n({\bf r};q) \right\}, \hspace{.5cm} q\in[0,1],   \label{geq:zeroth}
\end{eqnarray}
where $\psi_n^{(0)}({\bf r})$ is the initial guess of $\psi_n({\bf r})$, $\Psi_n({\bf r};q)$ and ${\cal E}_n(q) $ are the continuous mappings in $q\in[0,1]$ for $\psi_n({\bf r})$ and $E_n$, respectively.  Note that we have great freedom to choose the initial guess $\psi_n^{(0)}({\bf r})$  and the convergence-control parameter $c_0$.  When $q=0$, we have 
\begin{equation}
 \left( H_0 - E_n^b\right)  \left[ \Psi_n({\bf r};0)  -  \psi_n^{(0)}({\bf r}) \right] = 0, \label{geq:q=0}
\end{equation}
which gives, since $\left(H_0-E_n^b \right)$ is a linear operator, that  
\begin{equation}
\Psi_n({\bf r};0)  = \psi_n^{(0)}({\bf r}).  \label{Psi:E:q=0}
\end{equation}
When $q=1$,  since $c_0 \neq 0$,  Eq. (\ref{geq:zeroth}) is equivalent to the original equation (\ref{geq:original}), provided
\begin{equation}
\Psi_n({\bf r};1)  =     \psi_n({\bf r}),   \;\;\;  {\cal E}_n(1) = E_n. \label{Psi:E:q=1}
\end{equation}
Write $  E_n^{(0)}={\cal E}_n(0)$.   Then, as $q$ enlarges from 0 to 1,  $\Psi_n({\bf r};q) $ varies (or deforms) continuously from the known initial guess  $\psi_n^{(0)}({\bf r})$ to the unknown eigenfunction $\psi_n({\bf r})$, while ${\cal E}_n(q)$ changes  continuously  from $E_n^{(0)}$ to the unknown eigenvalue $E_n$, respectively.    This is the reason why  Eq. (\ref{geq:zeroth}) is called  the zeroth-order deformation equation.    

Note that the zeroth-order deformation equation  (\ref{geq:zeroth})  contains the so-called ``convergence control parameter'' $c_0$,  which has no physical meaning so that we have great freedom to choose its value.  Thus,  both of  $\Psi_n({\bf r};q) $  and  ${\cal E}_n(q)$  are dependent upon $c_0$, too.     Assume that the convergence-control parameter  $c_0$ is so properly chosen that the Maclaurin series 
\begin{eqnarray}
\Psi_n({\bf r};q)  &=&  \psi_n^{(0)}({\bf r}) +\sum_{k=1}^{+\infty}  \psi_n^{(k)}({\bf r}) \; q^k,   \label{series:Psi}   \\   
{\cal E}_n(q) &=& E_n^{(0)} +\sum_{k=1}^{+\infty} E_n^{(k)} \; q^k ,  \label{series:En}
\end{eqnarray}
exist and besides are convergent at $q=1$, where
\[    \psi_n^{(k)}({\bf r}) = \frac{1}{k!} \left.   \frac{\partial^k \Psi_n({\bf r};q) }{\partial q^k} \right|_{q=0},  \;\;\; E_n^{(k)} =  \frac{1}{k!} \left.   \frac{d^k {\cal E}_n(q) }{d q^k} \right|_{q=0} .  \]
Then, according to (\ref{Psi:E:q=1}), we have the homotopy series solution
\begin{eqnarray}
\psi_n({\bf r})  &=&  \psi_n^{(0)}({\bf r}) +\sum_{k=1}^{+\infty}  \psi_n^{(k)}({\bf r}),  \label{series-solution:psi} \\
 E_n  &=& E_n^{(0)} +\sum_{k=1}^{+\infty} E_n^{(k)}.  \label{series-solution:En}
\end{eqnarray}
The $M$th-order approximation of $\psi_n({\bf r})$ and $E_n$ read
\begin{eqnarray}
\hat{\psi}_n({\bf r})  &\approx&  \psi_n^{(0)}({\bf r}) +\sum_{k=1}^{M}  \psi_n^{(k)}({\bf r}),  \label{series-solution:psi:Mth} \\
\hat{E}_n  &\approx& E_n^{(0)} +\sum_{k=1}^{M} E_n^{(k)}.  \label{series-solution:En:Mth}
\end{eqnarray}
The accuracy of the approximation is measured by the residual error square of the original Schr\"{o}dinger equation, i.e.  
\begin{equation}  
\tilde{\Delta}_M^{RES} = \left( H\hat{\psi}_n -\hat{E}_n \hat{\psi}_n , H\hat{\psi}_n^* -\hat{E}_n \hat{\psi}_n^* \right).   \label{def:RES}
\end{equation}

Substituting the Maclaurin series (\ref{series:Psi}) and (\ref{series:En}) into the zeroth-order deformation equation (\ref{geq:zeroth}) and equating the like-power of $q$,   we have the first-order deformation equation
\begin{equation}
 \left( H_0 - E_n^b\right)  \psi_n^{(1)} ({\bf r}) = c_0  \left[ H \psi_n^{(0)} ({\bf r}) - E_n^{(0)} \psi_n^{(0)} ({\bf r})\right]  = c_0  R_{n,0}({\bf r}) \label{geq:1st}
\end{equation}
and the high-order deformation equation
\begin{equation}
 \left( H_0 - E_n^b\right) \left[  \psi_n^{(m)} ({\bf r})-\psi_n^{(m-1)}({\bf r}) \right] = c_0 \; R_{n,m-1}({\bf r}),  \hspace{0.5cm} m\geq 2,  \label{geq:mth}
\end{equation}
where 
\begin{eqnarray}
R_{n,k}({\bf r}) &=&  \frac{1}{k!}  \left. \frac{\partial^k 
 \left\{  H \Psi_n({\bf r};q)  -  {\cal E}_n(q) \Psi_n({\bf r};q) \right\}
}{\partial q^k} \right|_{q=0} \nn\\
&=&  H \psi_n^{(k)}({\bf r})-\sum_{j=0}^{k}E_n^{(j)}\psi_n^{(k-j)} ({\bf r}). \label{def:R[n,m]}
\end{eqnarray}
 
Writing
\begin{equation}   
 \psi_n^{(1)}({\bf r})   =   \sum_{m=0}^{N_s}  a_{n,m}^{(1)} \; \psi_m^{b}({\bf r})   \label{res:psi[0]}
 \end{equation}
and using (\ref{geq:known}),  the 1st-order deformation equation (\ref{geq:1st}) becomes 
\begin{equation}
 \sum_{m=0}^{N_s}  a_{n,m}^{(1)} \;\left( E_m^{b} -E_n^b\right)  \; \psi_m^{b}({\bf r}) =  c_0  \; R_{n,0} ({\bf r}) .
\end{equation}
Multiplying $\psi_j^{b}({\bf r})^*$ on both sides of the above equation and integrating in the whole domain, we have  
\begin{equation}
 \sum_{m=0}^{N_s}  a_{n,m}^{(1)} \;\left( E_m^{b} - E_n^b \right) \; \left(\psi_m^{b}({\bf r}), \psi_j^{b}({\bf r})^*\right) =  c_0  \left( R_{n,0}({\bf r}), \psi_j^{b}({\bf r})^*\right) .
\end{equation}
Since $\left(\psi_m^{b}({\bf r}), \psi_j^{b}({\bf r})^*\right) = \delta_{mj}$, we have 
\begin{equation}
\left( E_m^{b} - E_n^b \right)a_{n,m}^{(1)} = c_0   \;  \Delta^{n,m}_0.   \label{eq:a[n,m]}
\end{equation}
where
\begin{equation}
\Delta^{n,m}_0 =  \left( R_{n,0}({\bf r}), \psi_m^{b}({\bf r})^*\right) = \int  \left[  R_{n,0}({\bf r}) \; \psi_m^{b} ({\bf r})^*  \right] d\Omega.
\end{equation}
So, in case of $m \neq n$,   we have
\begin{equation}
a_{n,m}^{(1)} = c_0   \; \left(  \frac{\Delta^{n,m}_0}{E_m^{b} - E_n^b } \right),  \;\; \;\; m \neq n.     \label{def:a[n,m]}
\end{equation}
However, in case of $m=n$, for {\em arbitrary}  finite value of $a_{n,n}^{(1)}$,   we always have 
\begin{equation}
\Delta^{n,n}_0 = a_{n,n}^{(1)} \left( E_n^b -E_n^b\right) = 0, 
\end{equation}
say, 
\begin{equation}
\left( H \psi_n^{(0)}({\bf r}),\psi_n^b({\bf r})^*  \right) - E_n^{(0)} \left( \psi_n^{(0)}({\bf r}) , \psi_n^b({\bf r})^* \right) = 0,
\end{equation}
which gives 
\begin{equation}
E_n^{(0)}  = \frac{\left( H \psi_n^{(0)}({\bf r}),\psi_n^b({\bf r})^*  \right) }{ \left( \psi_n^{(0)}({\bf r}), \psi_n^b({\bf r})^* \right) }.  \label{def:En[0]}
\end{equation}
Thus, we have the solution
\begin{equation}
\psi_n^{(1)}({\bf r})   =  a_{n,n}^{(1)} \; \psi_n^b({\bf r}) + c_0 \sum_{m=0, 
m\neq n}^{N_s} \left(  \frac{\Delta^{n,m}_0}{E_m^{b}-E_n^b} \right)\; \psi_m^{b}({\bf r}),  \label{res:psi[1]}
\end{equation}
where the cofficient $a_{n,n}^{(1)} $ is unknown.   

Similarly, writing
\[    \psi_n^{(k)}({\bf r}) - \psi_n^{(k-1)}({\bf r})   =   \sum_{m=0}^{N_s}  a_{n,m}^{(k)} \; \psi_m^{b}({\bf r}), \hspace{1.0cm} k\geq 2  \]
and using (\ref{geq:known}),  we have
\begin{equation}
a_{n,m}^{(k)} = c_0 \left(\frac{\Delta^{n,m}_{k-1}}{E_m^{b}-E_n^b}\right), \;\;\; m\neq n,  \label{def:a[n,j][k]}
\end{equation}
where
\begin{equation}
\Delta^{n,m}_{j} =  \left( R_{n,j}({\bf r}), \psi_m^{b}({\bf r})^*\right) = \int  \left[  R_{n,j}({\bf r}) \; \psi_m^{b} ({\bf r})^* \right] d\Omega
\end{equation}
is the projection of $R_{n,j}({\bf r}) $ on $\psi_m^{b}({\bf r})$,   
so that
\begin{equation}
\psi_n^{(k)}({\bf r})   =  \psi_n^{(k-1)}({\bf r})+ a_{n,n}^{(k)} \; \psi_n^b({\bf r}) + c_0 \sum_{m=0, m\neq n}^{N_s} \left(  \frac{\Delta^{n,m}_{k-1}}{E_m^{b}-E_n^b} \right)\; \psi_m^{b}({\bf r}), \hspace{1.0cm} k > 1,  \label{res:psi[k]}
\end{equation}
where the coefficient $a_{n,n}^{(k)} $ is unknown.  

Similarly, $E_n^{(k)}$ is determined by the equation 
\begin{equation}  
 \Delta^{n,m}_{k} = 0, 
\end{equation}
say,
\begin{equation}
\left( H\psi_n^{(k)}({\bf r}) -\sum_{j=0}^{k-1} E_n^{(j)} \psi_n^{(k-j)}({\bf r}), \psi_n^b({\bf r})^* \right) - E_n^{(k)} \left(\psi_n^{(0)}({\bf r}),\psi_n^b({\bf r})^* \right) = 0,
\end{equation}
which gives
\begin{equation}
E_n^{(k)} = \frac{\left( F_{n,k}({\bf r}), \psi_n^b({\bf r})^* \right)}{\left(\psi_n^{(0)}({\bf r}),\psi_n^b({\bf r})^* \right)}. \label{def:En[k]}
\end{equation}
where 
\begin{equation}
F_{n,k}({\bf r})= H\psi_n^{(k)}({\bf r}) -\sum_{j=0}^{k-1} E_n^{(j)} \psi_n^{(k-j)}({\bf r}).   
\end{equation}

Note that the coefficient $a_{n,n}^{(1)}$  in (\ref{res:psi[1]}) and  $a_{n,n}^{(k)}$  in (\ref{res:psi[k]})   are  unknown.  In the perturbation approach of quantum mechanics, they are {\em assumed} to be zero.   However,  seriously speaking,  they can be {\em arbitrary} in mathematics.  How to determine them ?   

Note that each value of $a_{n,n}^{(k)}$ corresponds to a residual error square (\ref{def:RES}) of the original Schr\"{o}dinger equation.    Obviously,  the optimal value of $a_{n,n}^{(k)}$ should give the minimum of the residual error square at the $k$th-order approximation.     Write
\begin{equation}
\hat{\psi}_n({\bf r})' = \sum_{j=0}^{k-1} \psi_n^{(j)}({\bf r})  +  \tilde{\psi}_n^{(k)}({\bf r}) ,
\end{equation}
where
\begin{equation}
\tilde{\psi}_n^{(k)}({\bf r})   =  \psi_n^{(k-1)}({\bf r})+ c_0 \sum_{m=0, m\neq n}^{N_s} \left(  \frac{\Delta^{n,m}_{k-1}}{E_m^{b}-E_n^b} \right)\; \psi_m^{b}({\bf r}).  
\end{equation} 
Then, the residual error square 
\begin{equation}
\tilde{\Delta}^{RES}_k=\left( \left( H-\hat{E}_n\right)\left( \hat{\psi}_n({\bf r})'  + a_{n,n}^{(k)} \; \psi_n^b \right), \left( H-\hat{E}_n\right)\left( \hat{\psi}_n({\bf r})'  + a_{n,n}^{(k)} \; \psi_n^b \right)^*\right)
\end{equation}
has the minimum when 
\[   \frac{d  \tilde{\Delta}^{RES}_k }{d  \, a_{n,n}^{(k)}}  = 0,\]
say,
\begin{equation}
\left( \left( H-\hat{E}_n\right)\left( \hat{\psi}_n({\bf r})'  + a_{n,n}^{(k)} \; \psi_n^b \right), \left( H-\hat{E}_n\right) (\psi_n^b)^* \right) = 0
\end{equation} 
which gives the optimal value
\begin{equation}
a_{n,n}^{(k)} = -\frac{\left(\left( H-\hat{E}_n\right) \hat{\psi}_n({\bf r})', \left( H-\hat{E}_n\right)  \psi_n^b({\bf r})^* \right) }{\left(  \left( H-\hat{E}_n\right)  \psi_n^b({\bf r}),  \left( H-\hat{E}_n\right)  \psi_n^b({\bf r})^*\right)}.  \label{res:a[n,n]}    
\end{equation}

First of all, we choose an initial guess $\psi_n^{(0)}({\bf r})$.  Note that we have great freedom to choose it.   For tiny  disturbance,   we can  simply choose  $\psi_n^{(0)}({\bf r}) = \psi_n^b({\bf r})$.    Then, we gain $E_n^{(0)}$ by means of (\ref{def:En[0]}) and  further obtain $\psi_n^{(1)}$ by (\ref{res:psi[1]}) and (\ref{res:a[n,n]} ) for the 1st-order deformation equation.  Thereafter, we gain $E_n^{(1)}$ by means of (\ref{def:En[k]}) and further  $\psi_n^{(2)}$ by (\ref{res:psi[k]}) and  (\ref{res:a[n,n]} ) for the 2nd-order deformation equation, and so on.  

Unlike perturbation methods, there exists the so-called ``convergence-control parameter'' $c_0$ in the frame of the HAM, which has no physical meanings so that we have great freedom to choose its value so as to guarantee the convergence of the solution series (\ref{series-solution:psi}) and (\ref{series-solution:En}).   As illustrated by Liao \cite{LiaoBook2003, LiaoBook2012}, there always exists such a finite interval of $c_0$, in which each value of $c_0$ can ensure that each solution series converges to the same result, although with different rates of convergence.  Besides, the optimal value of the convergence-control parameter $c_0$  corresponds to the minimum of the residual error square, as mentioned by Liao \cite{LIAO2010-CNSNS-B}.  

 Unlike perturbation method, the HAM provides us great freedom to choose the initial guess $\psi_n^{(0)}({\bf r})$.  Obviously, we can use a known  $M$th-order approximation $\hat{\psi}_n$ as a better initial guess $\psi_n^{(0)}$.  This gives us the $M$th-order iteration of HAM approach, which can greatly accelerate the convergence of solution series, as mentioned by Liao \cite{LiaoBook2003, LiaoBook2012} and illustrated below.         

\section{An illustrative example}

\begin{figure}[tbh]
\begin{center}
\includegraphics[scale=0.4]{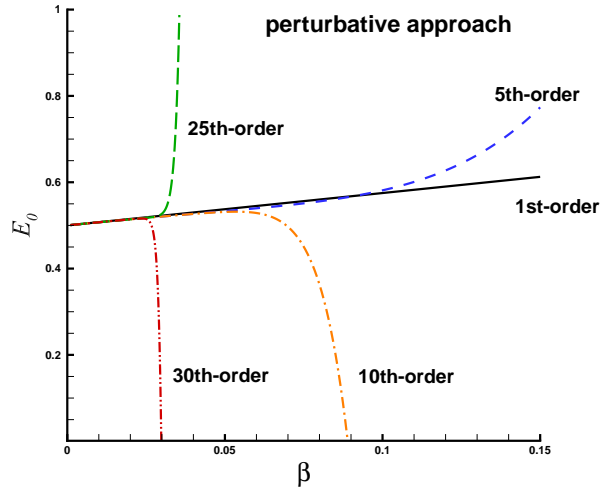}
\caption{The perturbative results of $E_0$ in (\ref{geq:exam1:original}) versus $\beta$  at the different order of approximation. Solid line: 1st-order; Dashed line: 5th-order; Dash-dotted line: 10th-order;  Long-dashed line: 25th-order; Dash-dot-dotted line: 30th-order.}
\label{Fig:E0:pert}
\end{center}
\end{figure}

\begin{figure}[tbh]
\begin{center}
\includegraphics[scale=0.4]{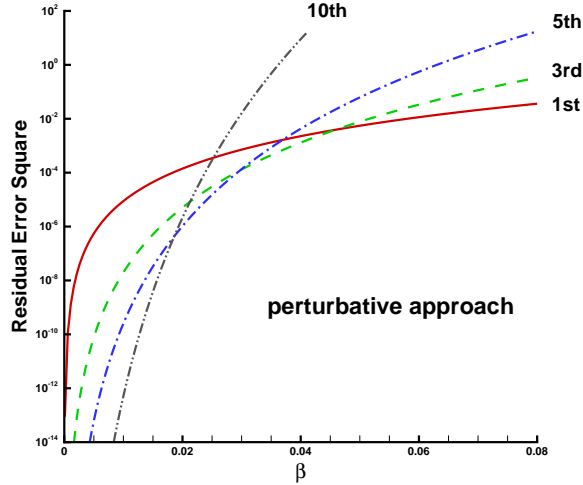}
\caption{The residual error squares of the perturbative results versus $\beta$ at the different order of approximations. Solid line: 1st-order; Dashed line: 3th-order; Dash-dotted line: 5th-order;  Dash-dot-dotted line: 10th-order.}
\label{Fig:RMS:pert}
\end{center}
\end{figure}

To show the validity of the HAM-based approach mentioned above, let us consider a one-dimensional nonlinear harmonic oscillator 
\begin{equation}
\tilde{H} {\tilde \psi}_n(x) =  \left[  -\frac{\hbar^2}{2m} \frac{d^2 }{d x^2}  + \frac{1}{2} m \omega^2 x^2 +\beta  \left(\frac{m^2 \omega^3}{\hbar} \right)x^4 \right] \tilde{\psi}_n(x)=\tilde{E}_n \; \tilde{\psi}_n(x).   \label{geq:exam1:original:0}
\end{equation}
Under the transformation 
\begin{eqnarray}
\xi = \sqrt{\frac{m \omega}{\hbar}} x,\;  \tilde{\psi}_n(x) = \left(\frac{m \omega}{\hbar}\right)^{1/4}\psi_n(\xi), \; \tilde{E}_n = \hbar \;  \omega \; E_n,
\end{eqnarray}
the dimensionless form of Eq. (\ref{geq:exam1:original:0}) reads
\begin{equation}
H \psi_n(\xi) = \left[ -\frac{1}{2} \frac{d^2}{d \xi^2} + \frac{1}{2} \xi^2 +\beta \xi^4 \right] \psi_n(\xi)= E_n \; \psi_n(\xi)  \label{geq:exam1:original}
\end{equation}
with an orthonormal basis
\begin{equation}
\psi_n^{b}(\xi) = \frac{1}{\sqrt[4]{\pi}\sqrt{ 2^n \; n!}} \; \tilde{H}_n(\xi) \; \exp\left( -\frac{\xi^2}{2}\right), \;\;  E_n^{b} = n + \frac{1}{2},
\end{equation}
satisfying 
\begin{equation}
H_0 \psi_n^{b}(\xi)  = E_n^{b} \;  \psi_n^{b}(\xi),
\end{equation}
where $E_n^{b} $ is the eigenfunction,  $\tilde{H}_n(\xi)$ is the $n$th Hermite polynomial in $\xi$, the Hamiltonian operators $H$ and $H_0$ are defined by
\begin{eqnarray}
H   &=&    -\frac{1}{2} \frac{d^2}{d \xi^2} + \frac{1}{2} \xi^2 + \beta \xi^4, \label{def:H} \\
H_0   &=&    -\frac{1}{2} \frac{d^2}{d \xi^2} + \frac{1}{2} \xi^2, \label{def:H0}
\end{eqnarray}
respectively.  

\subsection{Perturbative results}

The perturbative approach can be found in many textbooks of quantum mechanics.  For the sake of convenience, its basic ideas are briefly described in Appendix.  Here, we simply  give the perturbative result of $E_0$: 
\begin{equation}
E_0 = \frac{1}{2} +\frac{3}{4}\beta -\frac{21}{8}\beta^2+\frac{333}{16}\beta^3 -\frac{30885}{128}\beta^4+\frac{916731}{256}\beta^5-\frac{65518401}{1024}\beta^6 +\cdots \label{E0:pert}
\end{equation}

However, even for small values of $\beta$ such as $\beta=0.03$ and $\beta=0.05$, the perturbative results are unfortunately divergent, as shown in Tables~\ref{table:pert:beta0.03},  \ref{table:pert:beta0.05} and Fig.~\ref{Fig:E0:pert}.  At the 30th-order of approximation, the residual error squares of the perturbative series reach  $ 2.9 \times 10^{+19}$  in case of $\beta = 0.03$  and  $2.2 \times 10^{+41}$ in case of $\beta$ = 0.05, respectively.  As shown in Fig.~\ref{Fig:E0:pert}, although the 5th-order perturbative approximation of $E_0$ agrees well with the 10th-order approximation in $\beta\in[0,0.05]$,  the perturbation series of $E_0$ is actually divergent for $\beta\geq 0.02$.  This is very clear from  Fig.~\ref{Fig:RMS:pert}:  when $\beta>0.02$,   
the residual error squares of the perturbative results increase as the order of approximation enlarges.   Therefore, the traditional perturbative approach, which has been widely used in quantum mechanics, is  valid  only for a rather tiny disturbance indeed!   This greatly restricts the application of perturbation methods.  
        
\subsection{Results given by the HAM-based approach}

\begin{figure}[tb]
\begin{center}
\includegraphics[scale=0.4]{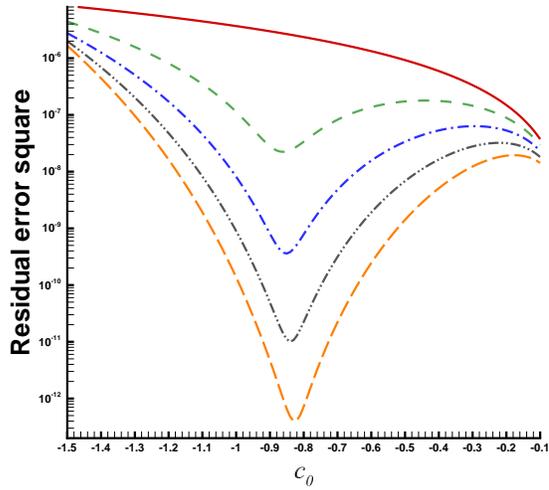}
\caption{Residual error square versus the convergence-control parameter $c_0$ in case of $\beta=0.01$ and $n=0$, given by means of the HAM-based approach using the initial guess $\psi_0^b({\bf r})$ and $N_s=24$. Solid line: 1st-order; Dashed line: 2nd-order; Dash-dotted line: 3rd-order; Dash-dot-dotted line: 4th-order; Long-dashed line: 5th-order.}
\label{Fig:RMS-beta0d01-HAM}
\end{center}
\end{figure}

\begin{figure}[htb]
\begin{center}
\includegraphics[scale=0.4]{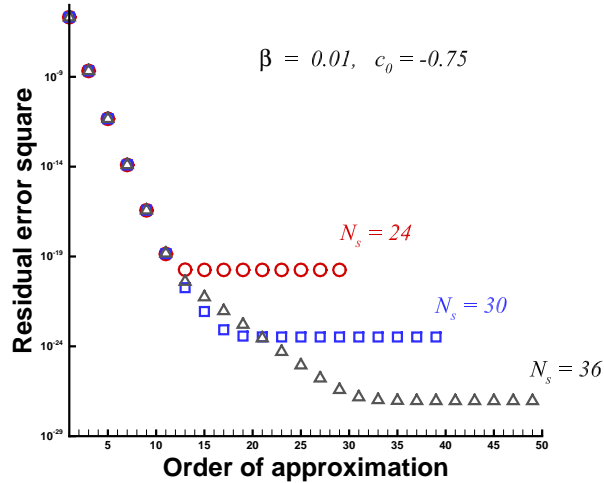}
\caption{Residual error square versus the order of approximation in case of $\beta=0.01$ and $n=0$, given by means of the HAM-based approach using the initial guess $\psi_0^b({\bf r})$ and $N_s=24, 30$ and 36, respectively. Circle: $N_s=24$; Square: $N_s=30$;  Delta: $N_s=36$. }
\label{Fig:RMS-order-beta0d01-HAM}
\end{center}
\end{figure}

However, unlike perturbative approach, the HAM contains the so-called ``convergence-control parameter'' $c_0$, which provides us a convenient way to guarantee the convergence of solution series, as shown below.    

To validate the HAM-baseed approach mentioned above, let us first consider the case with a small disturbance, i.e.  $\beta=1/100$.  We use the base  $\psi_0^b({\bf r})$  as  the initial guess of $\psi_0({\bf r})$ and choose $N_s=24$.  Note that, unlike the traditional perturbation approach,  the HAM approach contains the so-called ``convergence-control parameter'' $c_0$, so that all results including $E_0$, $\psi_0({\bf r})$ and the residual error square at different order of approximations are functions of $c_0$.   As shown in Fig.~\ref{Fig:RMS-beta0d01-HAM}, when $-1.5 < c_0<-0.1$,  the residual error squares continuously decrease as the order of approximation enlarges.  Besides,  the optimal value  of $c_0$ corresponds to  the minimum of the residual error square.  This is indeed true:  the convergent eigenfunction $\psi_0({\bf r})$ and eigenvalue  $E_0$ are gained by means of the HAM-based approach using  $c_0=-3/4$ and $N=24$, $30$ and 36, respectively, as shown in Fig.~\ref{Fig:RMS-order-beta0d01-HAM}.  Note that the ``final'' residual error square depends upon the truncation number $N_s$:  the larger the truncation number $N_s$, the smaller the ``final'' residual error square.  This is reasonable in mathematics, since larger truncation number $N_s$ should give better approximation.   By means of the HAM-based approach using $c_0=-3/4$ and $N_s=40$,  we gain  the convergent eigenvalue $E_0 = 0.50725620452460284095$ in accuracy of 20 digits, as shown in Table~\ref{exam1:beta0d01:E0},  which agrees well  with   its homotopy-Pad\'{e} approximation (see \cite{LiaoBook2003, LiaoBook2012}) in Table~\ref{exam1:beta0d01:E0:Pade}.    This illustrates the validity of the HAM-based approach for the time-independent Schr\"{o}dinger equation. 

\begin{figure}[tb]
\begin{center}
\includegraphics[scale=0.4]{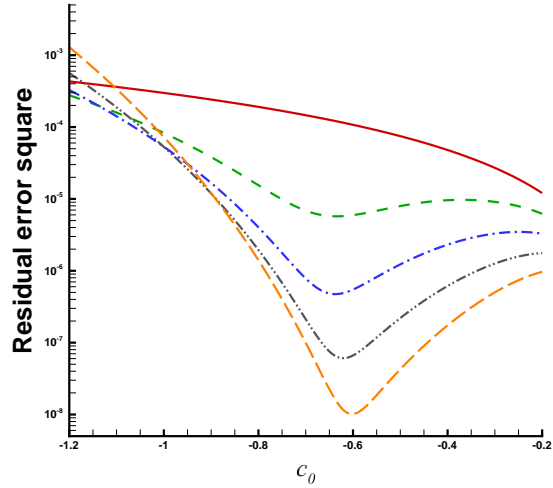}
\caption{The residual error square versus the convergence-control parameter $c_0$ in case of $\beta=0.03$ and $n=0$, given by means of the HAM-based approach using the initial guess $\psi_0^b({\bf r})$ and $N_s=40$ at the different orders of approximation. Solid line: 1st-order; Dashed line: 2nd-order; Dash-dotted line: 3rd-order; Dash-dot-dotted line: 4th-order; Long-dashed line: 5th-order.}
\label{Fig:RMS-beta0d03-HAM}
\end{center}
\end{figure}

\begin{figure}[htbp]
\begin{center}
\includegraphics[scale=0.4]{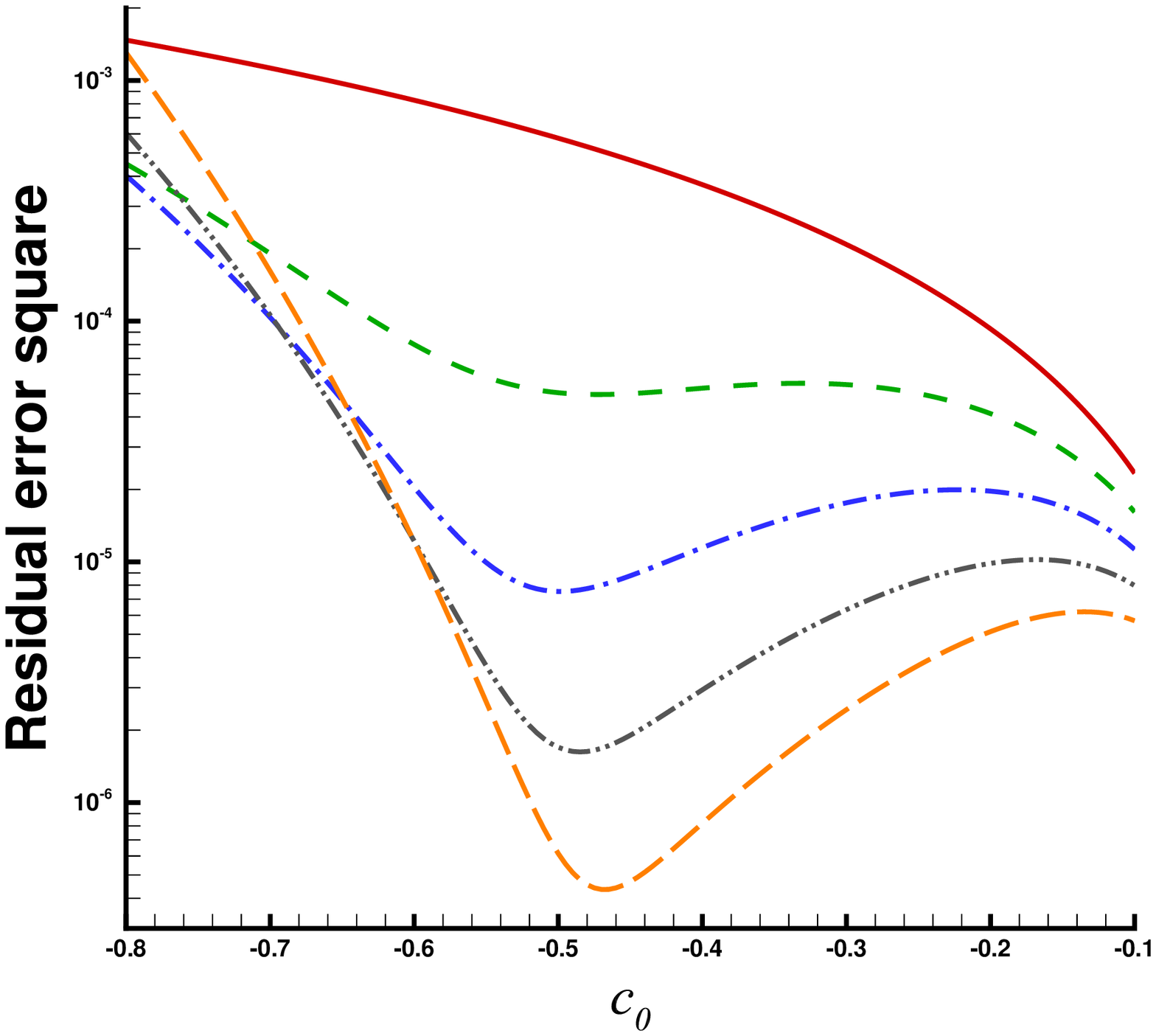}
\caption{Residual error square versus the convergence-control parameter $c_0$ in case of $\beta=0.05$ and $n=0$, given by means of the HAM-based approach using the initial guess $\psi_0^b({\bf r})$ and $N_s=40$ at the different orders of approximation. Solid line: 1st-order; Dashed line: 2nd-order; Dash-dotted line: 3rd-order; Dash-dot-dotted line: 4th-order; Long-dashed line: 5th-order.}
\label{Fig:RMS-beta0d05-HAM}
\end{center}
\end{figure}

\begin{figure}[htbp]
\begin{center}
\includegraphics[scale=0.4]{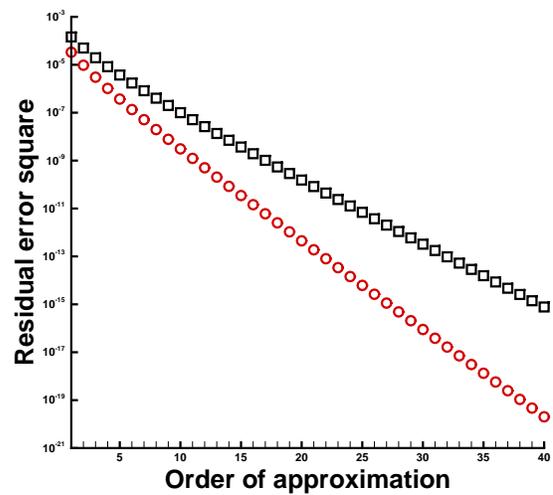}
\caption{Residual error square versus the order of approximation given by means of the HAM-based approach using the initial guess $\psi_0^b({\bf r})$ and $N_s=40$ in case of $\beta=0.03$, $n=0$, $c_0 = -1/3$  and  $\beta=0.05$, $n=0$, $c_0 = -1/4$, respectively.    Circle: $\beta=0.03$; Square: $\beta=0.05$. }
\label{Fig:RMS-order-beta0d03-5}
\end{center}
\end{figure}

\begin{figure}[htbp]
\begin{center}
\includegraphics[scale=0.4]{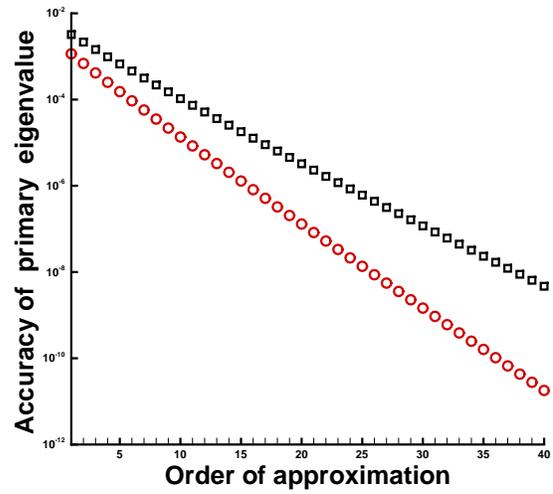}
\caption{The accuracy of the eigenvalue $E_0$ in case of  $\beta=0.03$ and $\beta=0.05$, given by means of the HAM-based approach  using the initial guess $\psi_0^b({\bf r})$, $N_s=40$  and  $c_0 = -1/3$ (when $\beta=0.03$) or $c_0 = -1/4$ (when $\beta=0.05$).   Circles; $\beta=0.03$: Square: $\beta=0.05$. }
\label{Fig:E0-accuracy}
\end{center}
\end{figure}

\begin{figure}[htbp]
\begin{center}
\includegraphics[scale=0.4]{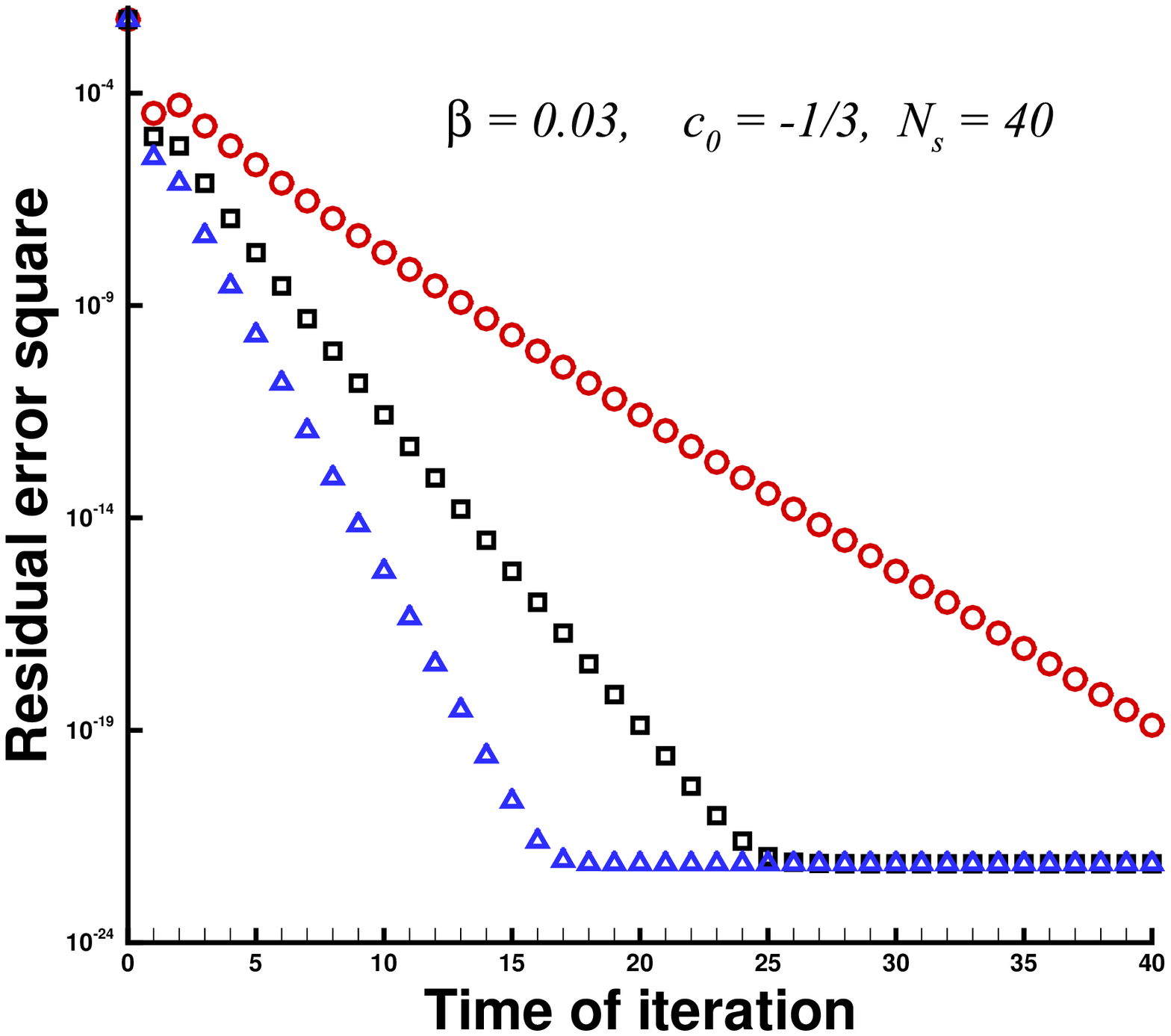}
\caption{Residual error square versus the times of iteration  in case of $\beta=0.03$ and $n=0$, given by means of the HAM-based approach using  $c_0 = -1/3$ and $N_s=40$.    Circle: 1st-order; Square: 2nd-order; Delta: 3rd-order. }
\label{Fig:RMS-iteration-beta0d03}
\end{center}
\end{figure}

\begin{figure}[htbp]
\begin{center}
\includegraphics[scale=0.4]{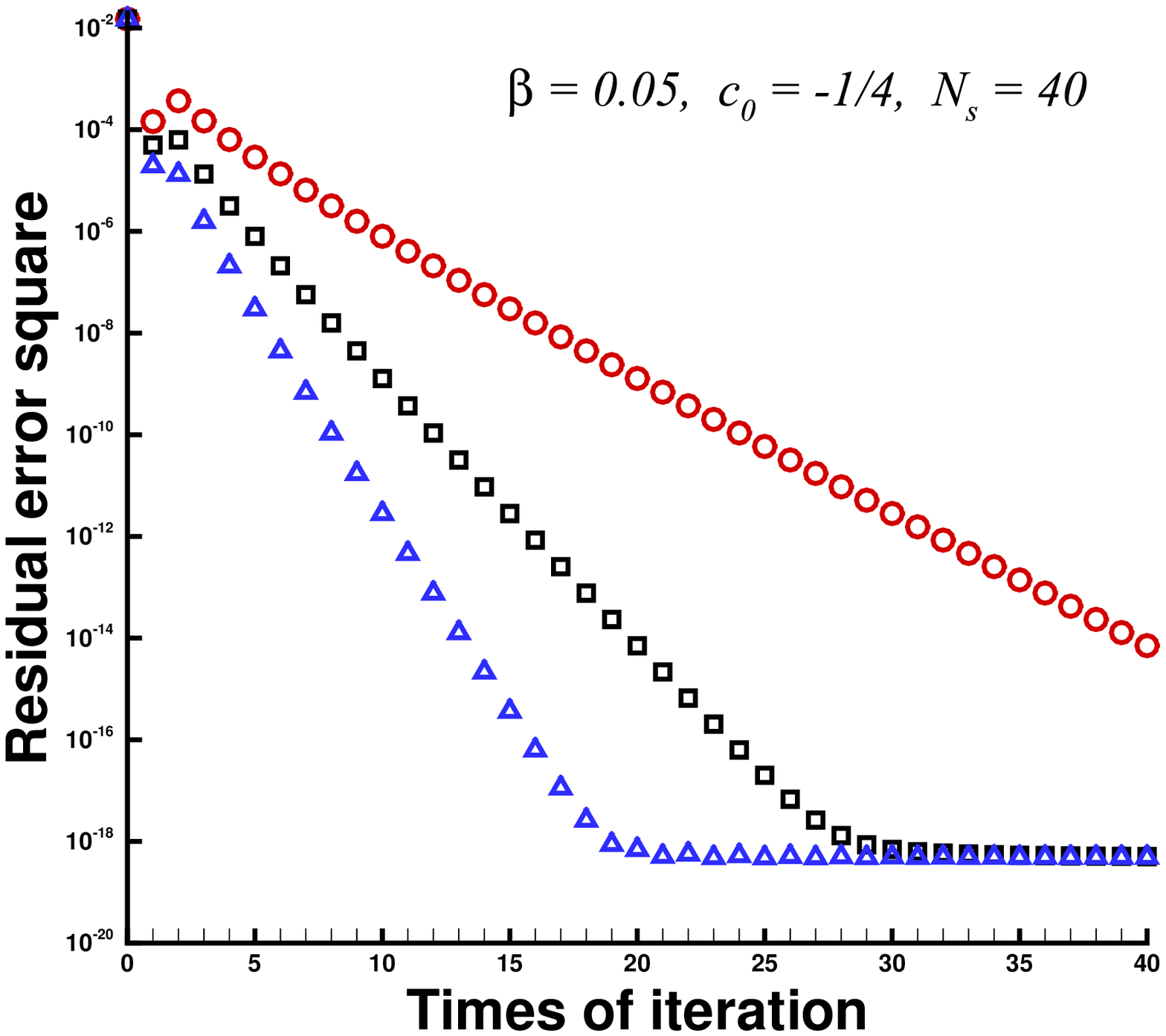}
\caption{Residual error square versus the times of iteration  in case of $\beta=0.05$ and $n=0$, given by means of the HAM-based approach using  $c_0 = -1/4$  and $N_s=40$.    Circle: 1st-order; Square: 2nd-order; Delta: 3rd-order. }
\label{Fig:RMS-iteration-beta0d05}
\end{center}
\end{figure}

\begin{figure}[htbp]
\begin{center}
\includegraphics[scale=0.4]{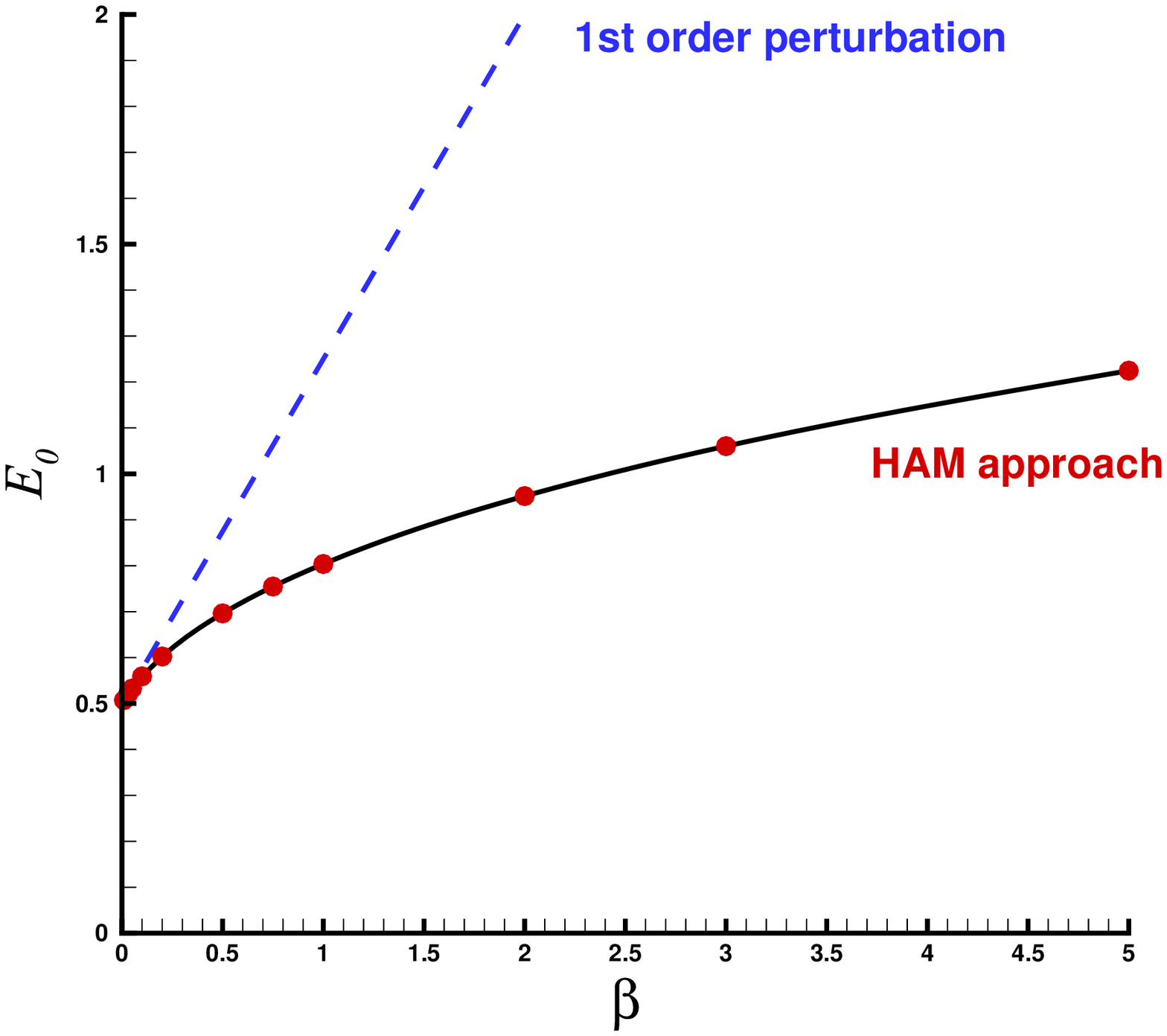}
\caption{The convergent results of the eigenvalue $E_0$ given by the HAM approach.  Solid line: HAM approach; Dashed line: 1st-order  perturbative approach. }
\label{Fig:E0:n0:HAM}
\end{center}

\begin{center}
\includegraphics[scale=0.4]{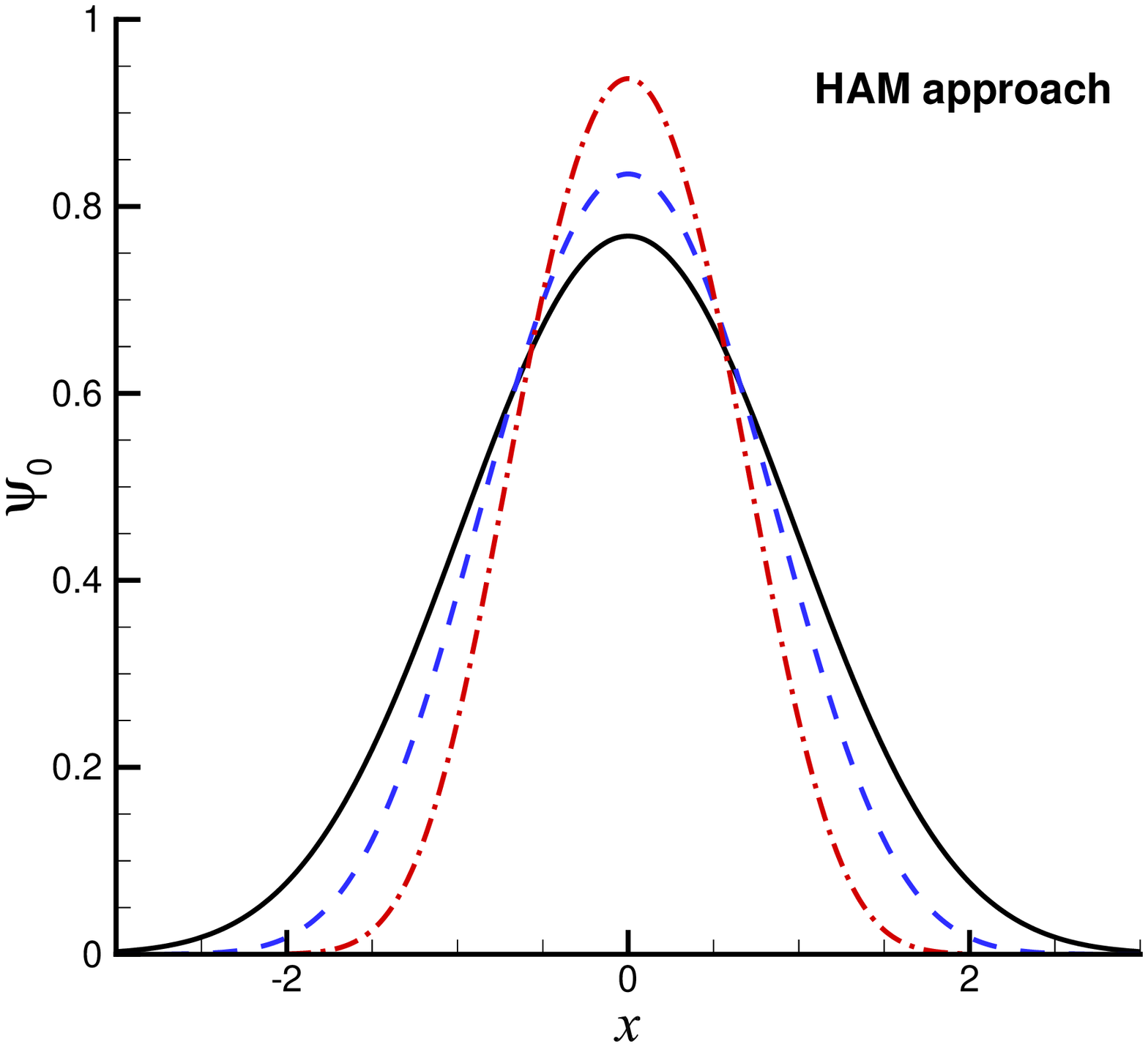}
\caption{The convergent results of the eigenfunction $\psi_0({\bf r})$ given by the HAM approach.  Solid line: $\beta=0.05$; Dashed line: $\beta=0.5$; Dash-dotted line: $\beta = 3$. }
\label{Fig:psi:n0:HAM}
\end{center}
\end{figure}

Note that  the perturbative series are divergent in case of $\beta=0.03$ and $\beta=0.05$, as shown in Figs.~\ref{Fig:E0:pert}, \ref{Fig:RMS:pert} and  Tables~\ref{table:pert:beta0.03}, \ref{table:pert:beta0.05}.    Fortunately, the HAM-based approach  contains  the  ``convergence-control parameter'' $c_0$, which has no physical meaning so that we have great freedom to choose it.   This provides us a convenient way to guarantee the convergence of solution series.   The residual error squares of the analytic approximations (versus $c_0$) given by the HAM-based approach using the initial guess $\psi_0^{(0)}({\bf r})$= $\psi_0^b({\bf r})$ and $N_s=40$ in case of $\beta=0.03$ and $\beta = 0.05$ are as shown in Figs.~\ref{Fig:RMS-beta0d03-HAM} and \ref{Fig:RMS-beta0d05-HAM}, respectively.   Note that  in each case there always exists a finite interval of $c_0$,  in which the residual error square decreases as the order of approximation enlarges.   This is indeed true:       
the convergent results are obtained by means of the HAM in case of $n=0$ and $\beta=0.03$ or $\beta=0.05$, using a proper ``convergence-control parameter'' $c_0=-2/5$ or $c_0=-1/4$, respectively, as shown in Tables~\ref{exam1:beta0d03:E0} to \ref{exam1:beta0d05:E0:Pade}.     In case of $\beta=-0.03$, our 40th-order approximation $E_0 = 0.52056172$ agrees  in accuracy of 8 digits with $E_0 = 0.52056171987300195300$ given by means of the homotopy-Pad\'{e} technique \cite{LiaoBook2003, LiaoBook2012}.  In case of $\beta=0.05$,  our 40th-order approximation $E_0 = 0.53264276$ agrees  in accuracy of 8 digits with $E_0 = 0.53264275477185884443$ given by means of the homotopy-Pad\'{e} technique \cite{LiaoBook2003, LiaoBook2012}.   With the increase of the approximation order,  the residual error squares decrease exponentially,   as shown in Figure~\ref{Fig:RMS-order-beta0d03-5}, while the accuracy of the eigenvalue $E_0$ increases exponentially, as shown in \ref{Fig:E0-accuracy}, respectively.    So, unlike perturbation approach that is {\em invalid} for $\beta>0.02$,  the HAM-based approach works well for larger disturbances $\beta=0.03$ and $\beta=0.05$.  These illustrate the validity of the HAM approach for the time-independent  Schr\"{o}dinger equations.

In addition, unlike perturbation method,  we have great freedom to choose the initial guess $\psi_n^{(0)}({\bf r})$ in the frame of the HAM to gain a $M$th-order approximation (\ref{series-solution:psi:Mth}).  Then, one can further use the known $M$th-order approximation as a new initial guess $\psi_n^{(0)}({\bf r})$ to gain a better $M$th-order approximation, and so on.   This provides us an iteration approach.    For example, in case of  $n=0$ and $\beta=0.03$ or $\beta=0.05$, we can first use the base $\psi_0^b({\bf r})$ as an initial guess of the unknown eigenfunction to gain a $M$th-order approximation  (\ref{series-solution:psi:Mth}) by means of the HAM-based approach with $N_s=40$ and $c_0=-1/3$ (when $\beta = 0.03$) or $c_0=-1/4$ (when $\beta=0.05$), and then use this $M$th-order approximation  as a new initial guess $\psi_n^{(0)}({\bf r})$ to further gain a better $M$th-order approximation (\ref{series-solution:psi:Mth})  by means of the HAM-based approach, where $M$=1,2,3.    As shown in Fig.~\ref{Fig:RMS-iteration-beta0d03},  the corresponding residual error squares decrease quickly, and besides the higher the order $M$ of approximation at each iteration,  the faster the iteration converges.  Note that, for the 2nd and 3rd-order iteration approach, the accuracy of approximate solution can not be heightened after some times of iteration, mainly due to the restriction of the truncation number $N_s$.  Note that all of these results given by the iterative HAM-based approaches agree quite well with the previous non-iterative HAM approach.   These illustrate the validity of the iterative HAM-based  approach mentioned in this paper.  Similarly,   by means of the iterative approach using a proper ``convergence-control parameter'' $c_0$ with a large enough truncation number $N_s$,  the convergent eigenfunction $\psi_0$ in case of $\beta=0.05$ can be further used as an initial guess $\psi_n^{(0)}({\bf r})$  to gain a convergent  eigenfunction $\psi_0({\bf r})$ and a convergent eigenvalue $E_0$ in case of $\beta=0.1$, and so on.  

In this way, we successfully obtain the convergent eigenvalue $E_0$ and eigenfunction $\psi_0({\bf r})$  for different values of $\beta\in[0,5]$,  as listed in Table~\ref{table:E0:Psi:exam1} and shown in  Figs.~\ref{Fig:E0:n0:HAM} and \ref{Fig:psi:n0:HAM}.

\section{Concluding remarks}

A new non-perturbative approach is proposed to solve time-independent Schr\"{o}dinger equations in  quantum  mechanics  and  chromodynamics (QCD).   It is based on the homotopy analysis method (HAM)  \cite{liao92-phd, liao97-nlm, LiaoBook2003, LiaoBook2012, liao04-AMC, LIAO2009-CNSNS-A,  LIAO2010-CNSNS-B, LIAO2010-CNSNS-A} developed by the author for highly nonlinear equations.   Unlike perturbative methods, this HAM-based approach has nothing to do with small/large physical parameters.  Besides,  convergent series solution can be obtained even if the disturbance is far from the known status.   A  nonlinear harmonic oscillator  is used as an example to illustrate the validity of this approach for disturbances that might be  hundreds times larger than the  possible  superior limit  of the perturbative approach.  This HAM-based approach could provide us rigorous  theoretical  results  in  quantum  mechanics  and  chromodynamics (QCD), which can be directly compared with experimental data.   Obviously, this is of great benefit not only for improving the accuracy of  experimental measurements but also for validating physical theories.

Finally, it should be emphasized that the basic ideas of the HAM-based approach have general meanings  and thus can be widely used to solve other kinds of equations in quantum  mechanics  and  chromodynamics (QCD) so as to gain rigorous, reliable, convergent eigenvalues and eigenfunctions.    The author would  highly suggest to recalculate  the  whole  systems of  the quantum mechanics and the chromodynamics (QCD), and then directly compare the convergent theoretical results with the corresponding experimental data.            

\section*{Acknowledgement}   This work is partly supported by National Natural Science Foundation of China (Approval No. 11432009 and 91752104).  Thanks to Dr. Theophanes E. Raptis in  University of Athens, Laboratory of Physical Chemistry,  Greece, for his suggestion on applying  the HAM to quantum mechanics and  chromodynamics (QCD).      

\section*{Appendix : Brief description of perturbation method}

The perturbative approach in quantum mechanics can be found in many textbooks \cite{DiracBook1958, GreinerBook2000, HamekaBook2004}.  Consider the Sch\"{o}dinger equation 
\begin{equation}
H \psi_n({\bf r}) = E_n \psi_n({\bf r}).  \label{geq:original:pert}
\end{equation}
Write
\[    H = H_0  + H',  \]
where $H, H_0$ are Hamiltonian operators, $H'$ is a small ``disturbance'' from $H_0$.  Assume that $\psi_n$ can be expressed by
\[   \psi_n({\bf r})  = \sum_{k=1}^{N_s} c_k \psi_k^{(0)}({\bf r}), \]
with
\begin{equation} 
  H_0 \psi_k^{(0)}({\bf r}) = E_k^{(0)} \psi_k^{(0)}({\bf r}),    \label{def:H0}
\end{equation}  
where $\psi_k^{(0)}({\bf r}) $ and $E_k^{(0)}$ are the eigenfunction and eigenvalue of the Hamiltonian operator $H_0$.     

Let $\lambda$ denote a small parameter and  write
\begin{eqnarray}
H &=& H_0 + \lambda H', \\
\psi_n({\bf r}) &=& \psi_n^{(0)} + \sum_{k=1}^{+\infty} \psi_n^{(k)}({\bf r}) \; \lambda^k, \\
E_n &=& E_n^{(0)} + \sum_{k=1}^{+\infty} E_n^{(k)} \; \lambda^k.
\end{eqnarray}
Substitute them into (\ref{geq:original:pert}) and equate the like-power of $\lambda$, we have the perturbation equations
\begin{eqnarray}
  \left( H_0 -E_n^{(0)}\right) \psi_n^{(0)} &=& 0, \label{geqLpert:0th} \\
  \left( H_0 -E_n^{(0)}\right) \psi_n^{(1)} &=&  E_n^{(1)} \psi_n^{(0)} - H' \psi_n^{(0)},  \label{geqLpert:1th} \\ 
  \left( H_0 -E_n^{(0)}\right) \psi_n^{(m)} &=&  E_n^{(m)} \psi_n^{(0)} - H' \psi_n^{(m-1)} +\sum_{k=1}^{m-1} E_n^{(k)}\psi_n^{(m-k)}, \;\;\; m\geq 2. \label{geqLpert:mth} 
\end{eqnarray}
According to (\ref{def:H0}), the zeroth-order perturbation equation (\ref{geqLpert:0th}) is automatically satisfied.  Multiplying $\left(\psi_n^{(0)}\right)^*$  on the both sides of (\ref{geqLpert:1th} ) and then integrating in the whole domain,    we have, since  $ \psi_n^{(0)}({\bf r}) $ is orthonormal and $H_0$ is a Hermite operator, that  
\begin{eqnarray}
E_n^{(1)} 
&=&  \left( \psi_n^{(0)}, H'\psi_n^{(0)}\right) =\tilde{\Delta}^{(0)}_{n,n}.
\end{eqnarray}
Write 
\[  \psi_n^{(m)} = \sum_{l \neq n}  a_{n,l}^{(m)} \; \psi_l^{(0)} \]
and substitute it into (\ref{geqLpert:1th}), we have 
\begin{equation}
\sum_{l \neq n}  a_{n,l}^{(1)} \; \left(  E_l^{(0)}-E_n^{(0)} \right) \psi_l^{(0)} =  E_n^{(1)} \psi_n^{(0)} - H' \psi_n^{(0)}.
\end{equation}  
Multiplying $\left(\psi_k^{(0)}\right)^*$  on the both sides of (\ref{geqLpert:1th} ) and then integrating in the whole domain,  we have in a similar way that 
\begin{equation}
a_{n,l}^{(1)} = \frac{\tilde{\Delta}_{l,n}^{(0)}}{E_n^{(0)}-E_l^{(0)}},
\end{equation} 
where
\begin{equation}
\tilde{\Delta}_{l,n}^{(0)} = \left( \psi_l^{(0)}, H'\psi_n^{(0)}\right).
\end{equation}

Similarly, we have for $m\geq 2$ that
\begin{eqnarray}
E_n^{(m)} &=& \tilde{\Delta}^{(m-1)}_{n,n} - \sum_{k=1}^{m-1} E_n^{(k)} a_{n,n}^{{(m-k)}}, \\
a_{n,l}^{(m)} &=& \frac{\tilde{\Delta}_{l,n}^{(m-1)}-\sum_{k=1}^{m} E_n^{(k)}a_{n,l}^{(m-k)}}{E_n^{(0)}-E_l^{(0)}},
\end{eqnarray}
where
\begin{equation}
\tilde{\Delta}_{l,n}^{(m-1)} = \left( \psi_l^{(0)}, H'\psi_n^{(m-1)}\right).
\end{equation}

The $M$th-order perturbation approximation reads 
\begin{eqnarray}
\psi_n({\bf r}) &\approx & \psi_n^{(0)} + \sum_{k=1}^{M} \psi_n^{(k)}({\bf r}),  \\
E_n &\approx & E_n^{(0)} + \sum_{k=1}^{M} E_n^{(k)}.
\end{eqnarray}



\bibliography{quantum}

\bibliographystyle{unsrt}


\begin{table}[htb]
\caption{The $m$th-order perturbative approximation of $E_0$ for Eq.~(\ref{geq:exam1:original})  and  the corresponding residual error square 
$\tilde{\Delta}^{RES}_m$ in case of $\beta=0.03$ by means of $N_s=40$. }
\begin{center}
\begin{tabular}{|c|c|c|} \hline\hline
\hspace{0.5cm} $m$th-order \hspace{0.5cm}	&	\hspace{1.5cm} $E_0$\hspace{1.5cm}	&	\hspace{0.5cm} residual error square 	\hspace{0.5cm}	\\ \hline 
1	&	0.5225	&	7.1 $\times 10^{-4}$ \\ 
3	&	0.520699	&	1.3 $\times 10^{-4}$ \\ 
5	&	0.520591	&	1.3 $\times 10^{-4}$ \\ 
10	&	0.520555	&	1.6 $\times 10^{-2}$ \\ 
15	&	0.520577	&	1.6	$\times 10^{+2}$ \\ 
20	&	0.520389	&	3.2 $\times 10^{+7}$ \\ 
25	&	0.526920	&	2.6 $\times 10^{+13}$ \\ 
30	&	-0.104234	&	2.9 $\times 10^{+19}$ \\ 
\hline
\end{tabular}
\end{center}
\label{table:pert:beta0.03}

\caption{The $m$th-order perturbative approximation of $E_0$ for Eq.~(\ref{geq:exam1:original})  and  the corresponding residual error square $\tilde{\Delta}^{RES}_m$ in case of $\beta=0.05$ by means of $N_s=40$ . }
\begin{center}
\begin{tabular}{|c|c|c|} \hline\hline
\hspace{0.5cm} $m$th-order \hspace{0.5cm}	&	\hspace{1.5cm} $E_0$\hspace{1.5cm}	&	\hspace{0.5cm} residual error square 	\hspace{0.5cm}	\\ \hline
1	&	0.5375	&	5.5 $\times 10^{-3}$ \\ 
3	&	0.533539	&	7.8 $\times 10^{-3}$ \\ 
5	&	0.533150	&	6.2 $\times 10^{-2}$ \\ 
10	&	0.531198	&	1.2 $\times 10^{+3}$ \\ 
15	&	0.572766	&	2.0 $\times 10^{+9}$ \\ 
18	&	-0.119387	&	5.0 $\times 10^{+13}$ \\ 
20	&	-4.95995	&	6.9$\times 10^{+16}$ \\
25	&	2549.9	&	7.1 $\times 10^{+26}$ \\
30	&	-3.16 $\times 10^6$	&	2.2 $\times 10^{+41}$ \\ \hline
\hline
\end{tabular}
\end{center}
\label{table:pert:beta0.05}
\end{table}%

\begin{table}[htb]
\caption{Approximations of $E_0$ and the residual error square  $\tilde{\Delta}^{RES}_m$  in case of  $\beta=1/100$ and $n=0$, given by means of the HAM-based approach using $N_s=40$, $c_0=-3/4$ and the initial guess $\psi_0^b({\bf r})$.}
\begin{center}
\begin{tabular}{|c|l|c|} \hline\hline
\hspace{0.5cm} $m$\hspace{0.5cm}  &  \hspace{1.5cm} $E_0$  \hspace{1.5cm}  	&	\hspace{0.5cm} residual error square 	\hspace{0.5cm} \\ \hline
1	&	0.5073031250				&	2.1  $\times 10^{-6}$  \\
2	&	0.5072656133				&	5.5  $\times 10^{-8}$  \\
3	&	0.5072581884				&	2.1  $\times 10^{-9}$  \\
4	&	0.5072566129				&	9.5  $\times 10^{-11}$  \\
8	&	0.5072562054				&	6.7  $\times 10^{-16}$  \\
12	&	0.50725620452				&	1.7  $\times 10^{-20}$  \\
16	&	0.5072562045246			&	3.8  $\times 10^{-22}$  \\
20	&	0.5072562045246028		&	3.6  $\times 10^{-23}$  \\
25	&	0.5072562045246028409		&	2.2  $\times 10^{-24}$  \\
30	&	0.50725620452460284095	&	1.3  $\times 10^{-25}$  \\
35	&	0.50725620452460284095	&	7.9  $\times 10^{-27}$  \\
40	&	0.50725620452460284095	&	4.9  $\times 10^{-28}$  \\
\hline\hline
\end{tabular}
\end{center}
\label{exam1:beta0d01:E0}
 \end{table}%

 \begin{table}[htb]
\caption{The $[m,m]$ homotopy-Pad\'{e} approximant of $E_0$ of (\ref{geq:exam1:original}) in case of  $\beta=1/100$ and $n=0$, given by means of the HAM-based approach using  $N_s=40$, $c_0=-3/4$ and the initial guess $\psi_0^b({\bf r})$.}
\begin{center}
\begin{tabular}{|c|l|} \hline\hline
\hspace{1.0cm} $m$  \hspace{1.0cm}   &	 \hspace{2.0cm}  $E_0$   \hspace{2.0cm}  \\ \hline 
2	&	0.50725620742     \\
4	&	0.507256204524	\\
6	&	0.507256204524602	\\
8	&	0.50725620452460284	\\
10	&	0.50725620452460284095	\\
12	&	0.50725620452460284095	\\ 
12	&	0.50725620452460284095	\\ 
12	&	0.50725620452460284095	\\ 
12	&	0.50725620452460284095	\\ 
20	&	0.50725620452460284095 \\
\hline\hline
\end{tabular}
\end{center}
\label{exam1:beta0d01:E0:Pade}
\end{table}%

\begin{table}[htb]
\caption{Approximations of $E_0$ and the residual error square  $\tilde{\Delta}^{RES}_m $  in case of  $\beta=3/100$ and $n=0$, given by means of the HAM-based approach using $N_s=40$, $c_0=-1/3$ and the initial guess $\psi_0^b({\bf r})$.}
\begin{center}
\begin{tabular}{|c|c|c|}  \hline\hline
\hspace{0.5cm} $m$\hspace{0.5cm}  &  \hspace{1.5cm} $E_0$  \hspace{1.5cm}  	&	\hspace{0.5cm} residual error square \hspace{0.5cm} \\ \hline
1	&	0.5217125	&	3.3 $\times 10^{-5}$  \\
3	&	0.52097595	&	3.0 $\times 10^{-6}$  \\
5	&	0.52071428	&	3.7 $\times 10^{-7}$  \\
10	&	0.52057524	&	3.1 $\times 10^{-9}$  \\
15	&	0.52056301	&	3.5	$\times 10^{-11}$  \\
20	&	0.52056185	&	4.3	$\times 10^{-13}$  \\
25	&	0.52056173	&	6.2	$\times 10^{-15}$  \\
30	&	0.52056172	&	8.9	$\times 10^{-17}$  \\
35	&	0.52056172	&	1.3	$\times 10^{-18}$  \\
40	&	0.52056172	&	2.0	$\times 10^{-20}$  \\
\hline\hline
\end{tabular}
\end{center}
\label{exam1:beta0d03:E0}
\end{table}%

\begin{table}[htb]
\caption{The $[m,m]$ homotopy-Pad\'{e} approximant of $E_0$ of (\ref{geq:exam1:original}) in case of  $\beta=3/100$ and $n=0$, given by means of the HAM using  $N_s=40$, $c_0=-1/3$ and the initial guess $\psi_0^b({\bf r})$.}
\begin{center}
\begin{tabular}{|c|l|}  \hline\hline
\hspace{1.0cm} $m$  \hspace{1.0cm}   &	 \hspace{2.0cm}  $E_0$   \hspace{2.0cm}  \\ \hline 
2	&	0.520562	\\ 
4	&	0.5205617	\\
6	&	0.5205617198	\\
8	&	0.520561719873	\\
10	&	0.52056171987300	\\
12	&	0.520561719873002	\\
14	&	0.52056171987300195	\\
16	&	0.52056171987300195300	\\
18	&	0.52056171987300195300	\\
20	&	0.52056171987300195300	\\
\hline\hline
\end{tabular}
\end{center}
\label{exam1:beta0d03:E0:Pade}
\end{table}%

\begin{table}[htb]
\caption{Approximations of $E_0$ and the residual error square  $\tilde{\Delta}^{RES}_m$  in case of  $\beta=1/20$ and $n=0$, given by means of the HAM-based approach using $N_s=40$, $c_0=-1/4$ and the initial guess $\psi_0^b({\bf r})$.}
\begin{center}
\begin{tabular}{|c|c|c|}  \hline\hline
\hspace{0.5cm} $m$\hspace{0.5cm}  &  \hspace{1.5cm} $E_0$  \hspace{1.5cm}  	&	\hspace{0.5cm} residual error square 	\hspace{0.5cm} \\ \hline
1	&	0.53585938	&	1.4 $\times 10^{-4}$  \\
3	&	0.53408899	&	2.0 $\times 10^{-5}$  \\
5	&	0.53331047	&	3.7 $\times 10^{-6}$  \\
10	&	0.53274784	&	1.0 $\times 10^{-7}$  \\
15	&	0.53266070	&	3.7 $\times 10^{-9}$  \\
20	&	0.53264599	&	1.5 $\times 10^{-10}$  \\
25	&	0.53264336	&	6.9 $\times 10^{-12}$  \\
30	&	0.53264287	&	3.2 $\times 10^{-13}$  \\
35	&	0.53264278	&	1.6 $\times 10^{-14}$  \\
40	&	0.53264276	&	7.8 $\times 10^{-16}$  \\
\hline\hline
\end{tabular}
\end{center}
\label{exam1:beta0d05:E0}
\end{table}%

\begin{table}[htb]
\caption{The $[m,m]$ homotopy-Pad\'{e} approximant of $E_0$ of (\ref{geq:exam1:original}) in case of  $\beta=1/20$ and $n=0$, given by means of the HAM-based approach using  $N_s=40$, $c_0=-1/4$ and the initial guess $\psi_0^b({\bf r})$.}
\begin{center}
\begin{tabular}{|c|l|}  \hline\hline
\hspace{1.0cm} $m$  \hspace{1.0cm}   &	 \hspace{2.0cm}  $E_0$   \hspace{2.0cm}  \\ \hline 
2	&	0.5326	\\
4	&	0.532642	\\
6	&	0.53264275	\\
8	&	0.532642754	\\
10	&	0.53264275477	\\
12	&	0.532642754772	\\
14	&	0.5326427547718	\\
16	&	0.53264275477185884	\\
18	&	0.53264275477185884443	\\
20	&	0.53264275477185884443	\\
\hline\hline
\end{tabular}
\end{center}
\label{exam1:beta0d05:E0:Pade}
\end{table}%

\begin{table}[htb]
\caption{Convergent results of the eigenvalue $E_0$ of Eq. (\ref{geq:exam1:original}) versus $\beta$, given by means of the HAM-based approach. }
\begin{center}
\begin{tabular}{|c|c|c|c|c|} \hline\hline
\hspace{0.75cm} $\beta$ 	\hspace{0.75cm} &	\hspace{0.75cm} $E_0$	\hspace{0.75cm} &	\hspace{0.75cm}$c_0$\hspace{0.75cm}	&\hspace{0.75cm}	$N_s$	\hspace{0.75cm}  
 \\  \hline
0.01	&	0.507256	&	-1		&	40		
\\ \hline
0.03	&	0.520562	&	-0.4		&	40		
\\ \hline
0.05	&	0.532643	&	-0.25		&	40	 	
\\ \hline
0.1	&	0.559146	&	-0.1		&	40	 
\\ \hline
0.2	&	0.602405	&	-0.05		&	40	
\\ \hline
0.5	&	0.696176	&	-0.01		&	55	
\\ \hline
0.75	&	0.754708	&	-0.005	&	60	
 \\ \hline
1	&	0.803771 	&	-0.002	&	70	
 \\ \hline
2	&	0.951569	&	-0.002	&	70	 
\\ \hline 
3	&	1.060271	&	-0.002	&	75	
\\ \hline 
5	&	1.224719	&	-0.001	&	80	
 \\ \hline 
\hline  
\end{tabular}
\end{center}
\label{table:E0:Psi:exam1}
\end{table}%

\end{document}